\newif\ifShowKeys
\ShowKeystrue
\ShowKeysfalse

\documentclass[letterpaper,10pt]{article}

\pdfoutput=1
\usepackage[no-natbib-sort]{my-jheppub}

\ifShowKeys \usepackage[notcite]{showkeys} \fi

\usepackage{amsmath, amssymb}

\usepackage{amsthm}

\usepackage{bm,environ,mathrsfs,array,arydshln,booktabs,float,slashed,hyperref}
\usepackage{tensor} 	% Ratcliffe package to write tensors
\usepackage{wick}
\usepackage[nodayofweek]{date time}

\usepackage{graphicx,epsfig}
\usepackage{epic}

\usepackage[dvipsnames]{xcolor}
\definecolor{maroon}{rgb}{0.8,0.3,0.}

\usepackage{aurical}
\usepackage[T1]{fontenc}

\usepackage{framed}
\definecolor{shadecolor}{RGB}{255, 230, 204}

\allowdisplaybreaks

\newcommand{\be}{\begin{equation}}
\newcommand{\ee}{\end{equation}}
\newcommand{\mc}{\mathcal }

\newcommand{\la}{\label}
\newcommand{\eps}{\varepsilon}

\newcommand{\tr}{\text{Tr} }
\newcommand{\N}{\mathcal{N}}

\newcommand{\norm}[1]{\mathop{:} #1 \mathop{:}}
\newcommand{\normtwo}[1]{\smash{{}^{\scalebox{0.6}{$\circ$}}_{\scalebox{0.6}{$\circ$}}}\,#1\,
\smash{{}^{\scalebox{0.6}{$\circ$}}_{\scalebox{0.6}{$\circ$}}}}

\newcommand{\ghat}{\left(\frac{g^{2}}{8\pi^{2}}\right)}

\makeatletter
\DeclareFontFamily{OMX}{MnSymbolE}{}
\DeclareSymbolFont{MnLargeSymbols}{OMX}{MnSymbolE}{m}{n}
\SetSymbolFont{MnLargeSymbols}{bold}{OMX}{MnSymbolE}{b}{n}
\DeclareFontShape{OMX}{MnSymbolE}{m}{n}{
    <-6>  MnSymbolE5
   <6-7>  MnSymbolE6
   <7-8>  MnSymbolE7
   <8-9>  MnSymbolE8
   <9-10> MnSymbolE9
  <10-12> MnSymbolE10
  <12->   MnSymbolE12
}{}
\DeclareFontShape{OMX}{MnSymbolE}{b}{n}{
    <-6>  MnSymbolE-Bold5
   <6-7>  MnSymbolE-Bold6
   <7-8>  MnSymbolE-Bold7
   <8-9>  MnSymbolE-Bold8
   <9-10> MnSymbolE-Bold9
  <10-12> MnSymbolE-Bold10
  <12->   MnSymbolE-Bold12
}{}

\let\llangle\@undefined
\let\rrangle\@undefined
\DeclareMathDelimiter{\llangle}{\mathopen}%
                     {MnLargeSymbols}{'164}{MnLargeSymbols}{'164}
\DeclareMathDelimiter{\rrangle}{\mathclose}%
                     {MnLargeSymbols}{'171}{MnLargeSymbols}{'171}
\makeatother

\title{Double scaling limit of $\N=2$ chiral correlators with Maldacena-Wilson loop }

\author[a]{Matteo Beccaria} 
\abstract{
We consider $\N=2$ conformal  QCD in four dimensions 
and the one-point correlator of a class of chiral primaries with the circular $\frac{1}{2}$-BPS
Maldacena-Wilson loop. We analyze a 
recently introduced double scaling limit where 
the gauge coupling is weak while the R-charge of the chiral primary $\Phi$ is large. In particular, 
we consider the case 
 $\Phi=(\tr\varphi^{2})^{n}$ , where $\varphi$ is the complex scalar in the vector multiplet.
The  correlator defines a non-trivial scaling function at fixed 
$\kappa = n\,g_{\rm YM}^{2}$ and large $n$ that may be studied by localization. For any gauge group $SU(N)$
we provide the analytic expression of the first correction $\sim \zeta(3)\,\kappa^{2}$ 
 and prove its universality. In the $SU(2)$ and $SU(3)$ theories we compute the scaling functions
at order $\mc O(\kappa^{6})$. Remarkably, in the $SU(2)$ case
the scaling function is equal to an analogous quantity 
describing the chiral 2-point functions $\langle\Phi\overline\Phi\rangle$ in the same large
R-charge limit. We conjecture that this $SU(2)$ scaling function is computed at all-orders by a
$\N=4$ SYM expectation value of a matrix model object characterizing the one-loop contribution to the 
4-sphere partition function. The conjecture provides an explicit series expansion for the scaling function
and is checked at order $\mc O(\kappa^{10})$ by showing agreement with  the available data  in the sector of chiral 
2-point functions.
}
\affiliation[a]{Dipartimento di Matematica e Fisica Ennio De Giorgi,\\
Universit\`a del Salento \& INFN, Via Arnesano, 73100 Lecce, Italy} 
\emailAdd{matteo.beccaria@le.infn.it} 

\begin{document}
%\date{\currenttime}
%\begin{flushright}\boxed{\small{\tt \today \ \ - \ \  \currenttime }}\end{flushright}
\maketitle

% ===============================================================
%_____ Main text  _________________________________________________________

\section{Introduction}

Recently, a certain interest has been devoted to the large R-charge limit of four dimensional $\N=2$
superconformal theories  \cite{Hellerman:2017sur,Hellerman:2018xpi}.
Besides, this limit may be conveniently 
combined with a weak  coupling expansion and tuned in order to 
provide a  non-trivial scaling behaviour. In this  mixed regime we can neglect
instanton contributions  while keeping some interesting (scaled) coupling dependence.
Typically, one has a vanishing Yang-Mills coupling 
$g_{\rm YM}\to 0$ while the R-charge 
grows like $1/g_{\rm YM}^{2}$.
Extremal correlators of chiral primaries in conformal $\N=2$ QCD
have been computed in such a double scaling limit  \cite{Bourget:2018obm}. If $\Phi_{n}$ is a chiral 
primary whose R-charge increases linearly with $n$, one may
consider the normalized ratio between the  2-point functions in the  $\N=2$ theory
and in the $\N=4$ SYM universal parent theory. This defines the scaling function for gauge 
group $SU(N)$
\be
\la{1.1}
F^{\Phi}(\kappa; N) = \lim_{n\to \infty}\left. \frac{\langle\Phi_{n}\overline\Phi_{n}\rangle^{\N=2}}
{\langle\Phi_{n}\overline\Phi_{n}\rangle^{\N=4}}\right|_{\kappa = \text{fixed}},
\ee
where $\kappa = n\,g^{2}_{\rm YM}$ is the fixed coupling at large R-charge.
 For certain classes
of chiral primaries $\Phi_{n}$, it is possible to compute the perturbative expansion of 
the function $F^{\Phi}$
by exploiting  the integrable structure of the $\N=2$ partition function. In the simplest setup,
$\Phi_{n}$ is the maximal multi-trace tower
$\Phi_{n} = \Omega_{n}\equiv (\tr\varphi^{2})^{n}$ where $\varphi $
is the complex scalar field belonging to the $\N=2$ vector multiplet. The 
 2-point functions $\langle \Omega_{n}\,\overline\Omega_{n}\rangle$
 are then captured by a Toda equation following from the four 
dimensional $tt^{*}$ equations \cite{Baggio:2014ioa}, {\em i.e.} the  counterpart of the topological 
anti-topological  fusion equations of 2d SCFTs \cite{Cecotti:1991me,Cecotti:1991vb}. 
By exploiting the Toda equation in order to  control the R-charge dependence, it is possible to 
compute the scaling function in (\ref{1.1})  at rather high perturbative order \cite{Beccaria:2018xxl}. 
This approach, based on decoupled semi-infinite Toda equations,
has been  proved to admit generalizations to broader classes of primaries and is believed to 
be a general feature of  Lagrangian $\N=2$ superconformal theories
\cite{Bourget:2018fhe}.

\medskip
In this paper, we reconsider the  double scaling limit  in (\ref{1.1}), but for a different class of 
correlators, {\em i.e.} the 1-point function $\langle\Phi_{n}\,W\rangle$
of large R-charge chiral primaries $\Phi_{n}$ in presence of a circular $\frac{1}{2}$-BPS Maldacena-Wilson loop $W$ \cite{Maldacena:1998im,Rey:1998ik}. In conformal SQCD, it is possible to compute such 
correlators by localization as thoroughly studied in \cite{Billo:2018oog}.
%for several choices of the 
%chiral primaries. 
Other applications of localization to Wilson loops in $\N=2$ superconformal theories may be found in 
\cite{Rey:2010ry,Passerini:2011fe,Russo:2013sba,Fiol:2015spa}.
For our purposes, we shall again be interested in the maximal multi-trace case, {\em i.e.}
$\Phi=\Omega_{n}$. 

\medskip
As is well known, the localization computation is based on the 
%the general correlator between a (scalar) chiral primary and the 
%$\frac{1}{2}$-BPS Wilson loop  may be  
%computed by localization in terms of the 
partition function of a suitable deformation 
of the $\N=2$ theory on $S^{4}$ \cite{Pestun:2016zxk,Nekrasov:2002qd,Nekrasov:2003rj}. 
However, the map from $S^{4}$ to flat space requires to disentangle a peculiar  operator mixing induced
by the conformal anomaly \cite{Gerchkovitz:2016gxx}. This is an annoying feature as far as 
the analysis of the $n\to\infty$ limit is concerned. Indeed,  the mixing structure  becomes more and more involved
with growing  $n$. In particular, one should need  results like those in \cite{Billo:2018oog}, but with a 
fully parametric dependence on $n$. Besides, as  in the study of the large R-charge limit of chiral 2-point functions, 
we want to work with a generic $SU(N)$ gauge group with (finite) fixed $N$. \footnote{
In particular, the $N\to \infty$ limit is not relevant to our study
and will be taken just to check our computation against some known results. }
The drawback is that 
finite $N$ results may display a deceiving complexity for large R-charge. 
For a discussion of what simplifications occur in the  mixing problem
at large $N$ see  \cite{Rodriguez-Gomez:2016cem,Baggio:2016skg,Rodriguez-Gomez:2016ijh,Pini:2017ouj}.

\medskip
To overcome these  difficulties, we exploit special features of the $\Omega_{n}$ operators.
For generic R-charge $n$ and gauge group $SU(N)$, 
we provide the solution to the mixing problem in the maximally supersymmetric $\N=4$ SYM
theory. Similarly, we give as well  exact 
expressions for the first genuine $\N=2$ correction $\sim \zeta(3)$. These findings are a 
useful guide
to study higher transcendentality contributions. 
Our results allows to consider the $n\to \infty$ limit of the ratio of 1-point functions
\be
\la{1.2}
G(\kappa; N) = \lim_{n\to \infty}\frac{\langle W\ \Omega_{n}\rangle^{\N=2}}
{\langle W\ \Omega_{n}\rangle^{\N=4}},
\ee
taken with fixed $\kappa = n\,g^{2}_{\rm YM}$ in the $SU(N)$ theory. The quantity in (\ref{1.2}) is the simplest
natural object to be studied in presence of the Wilson loop and corresponding to (\ref{1.1}).
Our analysis shows that the limiting scaling function $G(\kappa; N)$ is well defined and non-trivial.
In the $SU(2)$ theory, we check the remarkable equality $G(\kappa; N=2) = F(\kappa; N=2)$ at least at order 
$\mc O(\kappa^{6})$. This identity does not hold for $N>2$ as follows from a study of the 
 $G$ function in the  $SU(3)$ theory again at order $\mc O(\kappa^{6})$. 
 
\medskip
The case of $SU(2)$ is definitely special. Based on certain universality arguments we 
formulate a simple conjecture
for the all-order expansion of  $F(\kappa; 2)$  in terms of a certain $\N=4$
expectation value of the one-loop contribution to the $\N=2$ matrix model partition function. We have 
checked the conjecture at order $\mc O(\kappa^{10})$ by reproducing the results of \cite{Beccaria:2018xxl}.

\medskip
The plan of the paper is the following. In Section \ref{sec:largeR} we summarize recent developments
about the large R-charge double scaling limit in the sector of chiral 2-point correlators. 
In Section \ref{sec:wilson} we briefly set up the 
calculation with the Maldacena-Wilson loop
and recall the localization algorithm to map flat space correlators to definite matrix model
integrals. In Section \ref{sec:exact} we exploit some special features of the operators $\Omega_{n}$ to 
solve the associated mixing problem in the $\N=4$ theory, to compute the correlator with the Wilson loop
for a generic R-charge, and to evaluate the first correction $\sim \zeta(3)$ for a general gauge algebra rank.
Section \ref{sec:large} is devoted to a focused   study of the large R-charge limit. We first discuss in a rigorous 
way the universality of the $\zeta(3)$ correction. Next, simple educated assumptions allow us to compute 
the scaling function $G(\kappa; N)$ at sixth order in $\kappa$
for the $SU(2)$ and $SU(3)$ theories.  In the final 
Section \ref{sec:all} we present a conjecture for the all-order expression of the scaling function in the 
$SU(2)$ theory based on the identification of a particular universal and natural object in the $\N=2$ matrix model
which turns out to encode the scaling function itself.

\section{Large R-charge double scaling in  $\N=2$ conformal SQCD}
\la{sec:largeR}

We work in flat 4d space and consider $\N=2$ conformal SQCD, 
{\em i.e.} SYM with gauge group $SU(N)$ and $2N$
fundamental hypermultiplets. Chiral primary operators are primaries annihilated by half of the supercharges.
They have scaling dimension $\Delta$ and quantum numbers $(R, r)$ of the 
R-symmetry $SU(2)_{R}\times U(1)_{r}$ obeying $R=0$ and $\Delta=\frac{r}{2}$.
If $\varphi$ is the complex scalar field in the
vector multiplet, a generic  scalar chiral primary 
is labeled by  a vector of integers $\bm{n}=(n_{1}, \dots, n_{\ell})$
and reads \footnote{
See \cite{Buican:2014qla} for a recent discussion about general constraints on chiral primaries
in $\N=2$ superconformal theories prohibiting non-scalar chiral operators to a large extent.}
\be
\la{2.1}
\mc O_{\bm{n}}(x) = \prod_{k=1}^{\ell}\tr[\varphi(x)^{n_{k}}], \qquad \Delta = |n| = n_{1}+\dots+n_{\ell}.
\ee
Superconformal symmetry predicts the diagonal 2-point function
\be
\la{2.2}
\langle \mc O_{\bm{n}}(\infty) \,\overline{\mc O}_{\bm{m}}(0)\rangle = G_{\bm{n}, \bm{m}}\,
\delta_{|\bm{n}|, |\bm{m}|},
\ee
where $G_{\bm{n}, \bm{m}}$ is a function of the gauge coupling $g$. In the recent papers 
\cite{Bourget:2018obm,Beccaria:2018xxl} a special role has been played by the operators
\be
\la{2.3}
\Omega_{n} \equiv \mc O_{\underbrace{\scriptstyle 2, \dots, 2}_{n}} = (\tr\varphi^{2})^{n},
\ee
and the associated 2-point function coefficients
$G_{2n} = \langle (\tr\varphi^{2})^{n}(\tr\overline\varphi^{2})^{n}\rangle$ have been computed.
The functions $G_{2n}$  are captured by a semi-infinite Toda equation 
\cite{Gerchkovitz:2016gxx,Bourget:2018obm,Bourget:2018fhe}
that allows to compute them, after normalization to  their $\N=4$ SYM value, 
in the large R-charge limit  \footnote{Contrary to \cite{Bourget:2018obm},  here 
we denote  by $\kappa$ the large $n$ fixed coupling to 
avoid confusion with the more conventional 't~Hooft coupling $\lambda$.}
\be
\la{2.4}
n\to \infty, \qquad g\to 0, \qquad \kappa=n\,g^{2}=\text{fixed}.
\ee
One finds the expansion
\begin{align}
\la{2.5}
\log F(\kappa; N) &= \lim_{n\to \infty}\log \frac{G_{2n}^{\N=2}}{G_{2n}^{\N=4}} = 
 \sum_{\ell=2}^\infty \left(\frac{\kappa}{8\pi^2}\right)^\ell\,\sum_{\bm{s}} 
c^{(\ell)}_{\bm s}(N)\,\zeta(\bm{s}),
\qquad \zeta(\bm{s}) = \zeta(s_1)\,\zeta(s_2)\,\dots,
\end{align}
where the coefficients $c^{(\ell)}_{\bm s}(N)$ have been computed at order $\mc O(\lambda^{10})$
in \cite{Beccaria:2018xxl}. The first cases are 
%\begin{align}
%c^{(2)}_{3}(N) &= -18,\notag \\
%c^{(3)}_{5}(N) &= \frac{200 (2 N^2-1)}{N (N^2+3)},\notag \\
%c^{(4)}_{7}(N) &= -\frac{1225 (8 N^6+4 N^4-3 N^2+3)}{N^2 (N^2+1) (N^2+3) (N^2+5)}, \notag \\
%c^{(5)}_{9}(N) &= \frac{10584 (26 N^8+28 N^6-3 N^4+6 N^2-9)}{N^3 (N^2+1) (N^2+3) \
%(N^2+5) (N^2+7)}, \notag \\
%c^{(6)}_{5^2}(N) &= \frac{184800 (N^2-4) (N^6-N^4-43 N^2-37)}{(N^2+1) (N^2+3)^2 \
%(N^2+5) (N^2+7) (N^2+9)},\notag \\
%c^{(6)}_{11}(N) &= -\frac{71148 (122 N^{10}+280 N^8+48 N^6-15 N^4+45)}{N^4 (N^2+1) \
%(N^2+3) (N^2+5) (N^2+7) (N^2+9)}.
%\end{align}
{\small
\begin{align}
c^{(2)}_{3}(N) &= -18, & c^{(5)}_{9}(N) &= \frac{10584 (26 N^8+28 N^6-3 N^4+6 N^2-9)}
{N^3 (N^2+1) (N^2+3) (N^2+5) (N^2+7)}, \notag \\
c^{(3)}_{5}(N) &= \frac{200 (2 N^2-1)}{N (N^2+3)}, & 
c^{(6)}_{5^2}(N) &= \frac{184800 (N^2-4) (N^6-N^4-43 N^2-37)}{(N^2+1) (N^2+3)^2 \
(N^2+5) (N^2+7) (N^2+9)},\notag \\
c^{(4)}_{7}(N) &= -\frac{1225 (8 N^6+4 N^4-3 N^2+3)}{N^2 (N^2+1) (N^2+3) (N^2+5)}, &
c^{(6)}_{11}(N) &= -\frac{71148 (122 N^{10}+280 N^8+48 N^6-15 N^4+45)}{N^4 (N^2+1) \
(N^2+3) (N^2+5) (N^2+7) (N^2+9)}.
\end{align}
}
It has been conjectured that all terms associated with multiple products of zeta functions do vanish for $N=2$. 
For this value, {\em i.e.} for the $SU(2)$ theory, the expansion (\ref{2.5}) 
reduces to 
\begin{align}
\la{2.7}
& \log  F(\kappa; 2) = -\frac{9\, \zeta (3)}{32 \,\pi ^4}\, \kappa ^2 +\frac{25\,  \zeta 
(5)}{128 \,\pi ^6}\,\kappa ^3-\frac{2205\,  \zeta (7)}{16384 \,\pi 
^8}\,\kappa ^4+\frac{3213\,  \zeta (9)}{32768 \,\pi ^{10}}\,\kappa ^5 -\frac{78771 \,
 \zeta (11)}{1048576 \,\pi ^{12}}\,\kappa ^6\notag \\
 &+\frac{250965 \,
\zeta (13)}{4194304 \,\pi ^{14}}\,\kappa ^7 -\frac{105424605  \,\zeta 
(15)}{2147483648 \,\pi ^{16}}\,\kappa ^8 +\frac{265525975 \, \zeta 
(17)}{6442450944 \,\pi ^{18}}\,\kappa ^9-\frac{12108123027  \,\zeta 
(19)}{343597383680 \,\pi ^{20}}\,\kappa ^{10}+\mc O(\kappa ^{11}).
\end{align}
This simple exponentiation property is not true for $N > 2$. For instance, for $N=3$
one finds
\begin{align}
\la{2.8}
\log & F(\kappa; 3) =-\frac{9  \,\zeta (3)}{32 \,\pi ^4}\,\kappa^{2}+\frac{425  \,\zeta 
(5)}{2304 \,\pi ^6}\,\kappa^{3}-\frac{17885  \,\zeta (7)}{147456 \,\pi 
^8}\,\kappa^{4}+\frac{5565  \,\zeta (9)}{65536 \,\pi ^{10}}\,\kappa^{5}\notag \\
& +
\bigg(\frac{1925 \,\zeta (5)^2}{14155776 \,\pi ^{12}}-\frac{2668897 \,\zeta 
(11)}{42467328 \,\pi ^{12}}\bigg)\,\kappa^{6}+\bigg(\frac{32984237 \,\zeta 
(13)}{679477248 \,\pi ^{14}}-\frac{5005 \,\zeta (5) \,\zeta (7)}{14155776 \
\,\pi ^{14}}\bigg)\,\kappa^{7}\notag \\
& + \bigg(\frac{35035 \,\zeta (7)^2}{150994944 \,\pi 
^{16}}+\frac{146575 \,\zeta (5) \,\zeta (9)}{402653184 \,\pi 
^{16}}-\frac{2245755655 \,\zeta (15)}{57982058496 \,\pi ^{16}}\bigg)\,\kappa^{8}\notag\\
&+ \bigg(-\frac{1519375 \,\zeta (5)^3}{1174136684544 \,\pi 
^{18}}-\frac{3488485 \,\zeta (7) \,\zeta (9)}{7247757312 \,\pi 
^{18}}-\frac{546184925 \,\zeta (5) \,\zeta (11)}{1565515579392 \,\pi 
^{18}}\notag \\
&+\frac{669686057755 \,\zeta (17)}{21134460321792 \,\pi 
^{18}}\bigg)\,\kappa^{9}+\bigg(\frac{8083075 \,\zeta (5)^2 \,\zeta 
(7)}{1565515579392 \,\pi ^{20}}+\frac{77643709 \,\zeta 
(9)^2}{309237645312 \,\pi ^{20}}\notag \\
& +\frac{2905703801 \,\zeta (7) \,\zeta 
(11)}{6262062317568 \,\pi ^{20}}+\frac{4074100745 \,\zeta (5) \,\zeta 
(13)}{12524124635136 \,\pi ^{20}}-\frac{29805018472801 \,\zeta 
(19)}{1127171217162240 \,\pi ^{20}}\bigg)\,\kappa^{10}+\mc O(\kappa ^{11}),
\end{align}
and, starting at order $\kappa^{6}$, products of zeta functions appear.
This seemingly technical or accidental feature will have a role in the following discussion.
A natural issue 
is than to explore the possibility of non-trivial scaling functions in other partially protected sectors. To this aim,
we shall consider here  chiral correlators of one primary with a Maldacena-Wilson loop. 

\section{Chiral 1-point function with $\frac{1}{2}$-BPS Wilson loop}
\la{sec:wilson}

We consider the $\frac{1}{2}$-BPS Maldacena-Wilson loop defined by \cite{Maldacena:1998im,Rey:1998ik}
\be
W = \frac{1}{N}\,\tr\,\text{P}\,\exp\bigg[g\,\int_{C} dt\,\bigg(i\,A\cdot \dot x + \tfrac{R}{\sqrt{2}}\,
(\varphi+\overline\varphi)\bigg)\bigg],
\ee
where $g$ is the gauge coupling, $C$ is a circle of radius $R$,  $\varphi$ is the complex scalar field in the
vector multiplet, and the trace is taken in the fundamental representation.

\noindent
Conformal invariance
fully constrains the position dependence of the 1-point function of the chiral primary operator
$\mc O_{\bm{n}}$ with $W$. For a loop placed at the origin, one has \cite{Billo:2018oog}
\be
\la{3.2}
\langle W\,\mc O_{\bm{n}}(x)\rangle = \frac{A_{\bm n}(g)}{(2\pi\,|x|_{C})^{|\bm{n}|}},
\ee
where $|x|_{C}$ is a suitable $SO(1,2)\times SO(3)$
distance between $x$ and the loop respecting the conformal subgroup
unbroken by the loop. All the remaining 
information about the 1-point function is  encoded in the coupling dependent normalization 
$A_{\bm n}(g)$ in (\ref{3.2}).

\subsection{Localization results}

As discussed in \cite{Billo:2018oog}, 
the function $A_{\bm n}(g)$ may be computed 
by  the same matrix model that appears in the partition function and 
 encoding the localization solution of the $\N=2$ theory
on $S^{4}$ with a specific (finite) $\Omega$-deformation
% _{\eps_{1}, \eps_{2}}$ deformation with $\eps_{1}=\eps_{2}$ equal to 
%the inverse radius of $S^{4}$
 \cite{Pestun:2016zxk,Nekrasov:2002qd,Nekrasov:2003rj}. 
Since we are interested in a weak-coupling expansion, we shall 
drop the instanton contribution. After this simplification, the sphere 
partition function is associated with a perturbed Gaussian matrix model.
Up to a $g$-independent normalization it reads
\be
\la{3.3}
Z_{S^{4}} = \left(\frac{g^{2}}{8\pi^{2}}\right)^{\frac{N^{2}-1}{2}}\,\int Da\,
e^{-\tr a^{2}-S_\text{int}(a)},
\ee
where $a=a^{\ell}T^{\ell}$ is a traceless $N\times N$ matrix, $Da = \prod_{\ell=1}^{N^{2}-1}
\frac{da^{\ell}}{\sqrt{2\pi}}$, and the non-Gaussian {\em interacting action} $S_{\rm int}(a)$ is 
an infinite series
\be
\la{3.4}
S_\text{int}(a) = \sum_{n=0}^{\infty}s_{n}(a)\,\zeta(2n+3)\,g^{2n+4},
\ee
where $s_{n}(a)$ are invariant functions of $a$. The first terms read explicitly
\be
\la{3.5}
S_\text{int}(a) = \frac{3\,\zeta(3)\,g^{4}}{(8\pi^{2})^{2}}\,(\tr a^{2})^{2}-
\frac{10\,\zeta(5)\,g^{6}}{3\,(8\pi^{2})^{3}}\bigg[3\,\tr a^{4}\,\tr a^{2}-2\,(\tr a^{3})^{2}\bigg]+\dots,
\ee
and  higher powers of $g$ are associated with higher transcendentality terms.
As usual, correlation functions are computed by
\be
\la{3.6}
\langle F(a) \rangle = \frac{\int Da\, F(a)\,e^{-\tr a^{2}-S_\text{int}(a)}}
{\int Da\, e^{-\tr a^{2}-S_\text{int}(a)}}.
\ee
The $\N=4$ SYM theory is obtained as a special limit where the interacting action is dropped 
and the instanton contribution -- already discarded at weak-coupling -- is also removed. In this 
limit, the partition function is computed by a very simple Gaussian model 
and correlators are obtained after full Wick 
contraction where the basic contraction is $\langle a^{\ell}a^{\ell'}\rangle = \delta^{\ell\ell'}$. In the
following 
we shall be interested in the multi-trace expectation values, cf. (\ref{2.1}),
\be
\la{3.7}
\langle \mc O_{\bm n}(a)\rangle, \quad \text{where}\ \
\mc O_{\bm n}(a) \stackrel{\text{def}}{=} \prod_{k=1}^{\ell}\tr a^{n_{k}}.
\ee
As discussed in  \cite{Billo:2018oog}
 it is possible to compute the function $A_{\bm n}(g)$ in (\ref{3.2}) by the following
prescription
\be
\la{3.8}
A_{\bm n}(g) = \langle W(a)\, \norm{\mc O_{\bm n}(a)}\rangle,\qquad 
W(a) = \frac{1}{N}\,\tr\,\exp\left(\frac{g}{\sqrt 2}\,a\right),
\ee
where $\norm{O_{\bm n}(a)}$ is obtained by Gram-Schmidt orthogonalization with respect 
to all operators with  dimension smaller than $|\bm n|$. 
 In \cite{Billo:2018oog}, several interesting results are obtained for the function $A_{\bm n}(g)$ at specific 
 values of the multi-index $\bm n$, both at finite $N$ as well as  in the planar limit $N\to\infty$ with fixed 
 $Ng^{2}$. This is achieved by exploiting recursion relations with respect to $\bm{n}$ \cite{Billo:2017glv}. 
% This is quite 
%efficient for assigned $\bm n$, but in general it is less useful if one is interested in the analytical dependence on $\bm n$.
%
% 
% the aforementioned recursion relations obeyed by 
%the expectation values in (\ref{3.7}). 
In general, a  potential unavoidable problem is that it may be difficult to obtain  
results parametric in $\bm n$, whereas this is essential for our purposes. This is a difficulty already
showing up 
in the $\N=4$ theory due to the complication of the Gram-Schmidt procedure required to construct normal
ordering. In the $\N=2$ theory, complications are worse due to the $g$-dependence introduced by the 
interaction term (\ref{3.4}).

In the next Section we shall consider the special class of operators
$\Omega_{n}$ in (\ref{2.3}), show how to solve the above problems,  
and derive a set of results depending parametrically in $n$. This will be 
the starting point to discuss the large R-charge limit $n\to \infty$ in Section \ref{sec:large}.

\paragraph{Remark} Concerning notation, 
in the following we shall need to tell between quantities evaluated in the $\N=2$ or $\N=4$ 
theories. In such cases, we shall denote by $\langle \mc O\rangle$ and $\norm{\mc O}$
the expectation values and associated normal ordering in the $\N=4$ matrix model. Instead, we shall
denote by $\llangle \mc O\rrangle$ and $\normtwo{\mc O}$ the same quantities in the 
 $\N=2$ theory.
 
\section{The correlators $\langle W\,\Omega_{n}\rangle$}
\la{sec:exact}

As we explained in the Introduction, we are interested in the functions 
\be
\la{4.1}
A_{n}^{\N=2}(g) = \llangle W\,\normtwo{\Omega_{n}}\rrangle.
\ee
%that we may identify with a special case of (\ref{3.2})  or (\ref{3.8})  depending on the context. 
As we 
discussed, to some extent it is 
straighforward to evaluate the perturbative expansion of $A_{n}^{\N=2}$ at fixed $n$. However, 
our main concern is   the limit (\ref{2.4}) and the associated asymptotic ratio 
\be
\la{4.2}
\lim_{n\to\infty}\frac{A_{n}^{\N=2}}{A_{n}^{\N=4}} = 
\lim_{n\to\infty}\frac{\llangle W\,\normtwo{\Omega_{n}}\rrangle}{\langle W\,\norm{\Omega_{n}}\rangle},
\qquad \kappa = n\,g^{2} = \text{fixed}.
\ee
For this quantity a different approach is required. Notice that in the case of the 
chiral 2-point function, the existence of a Toda equation -- equivalent to the $tt^{*}$ equations --
has been crucial in this respect  \cite{Beccaria:2018xxl}.  In this section, we shall present results 
for $A_{n}(g)$ in the $\N=4$ theory. This is a piece entering the  ratio  (\ref{4.2}).
Next, we move to the $\N=2$ theory and compute the first non-trivial correction to $A_{n}^{\N=4}$,
{\em i.e.} the term proportional to $\zeta(3)$, cf. (\ref{3.5}). Although we are 
ultimately interested in the finite $N$ case, we shall check our results by matching
 general expressions valid in the planar limit that have been  
 conjectured in \cite{Billo:2018oog}. 
 
\subsection{Results in the $\N=4$ theory}

\paragraph{Normal ordering} 
Let $\mc O^{\Delta}$ be a basis for all multi-trace matrix model operators 
with dimension $\Delta$. A possible choice is $\mc O^{\Delta} = \{\mc O_{\bm n}(a), |\bm{n}|=\Delta\}$,
cf. (\ref{3.7}) . The Gram-Schmidt orthogonalization process amounts to write
\be
\la{4.3}
\norm{\Omega} = \Omega+\sum_{\Delta'<\Delta}\sum_{\Omega'\in\mc O^{\Delta'}}
c_{\Omega'}\,\Omega',
\ee
where the constants $c$ are determined by the condition $\langle\norm{\Omega}\Omega'\rangle=0$
for all $\Omega'\in\mc O^{\Delta'}$ with $\Delta'<\Delta$. A useful remark 
 is that  (\ref{4.3}) is equivalent
to the subtraction of all possible (partial) Wick pairings inside $\Omega$ \cite{Billo:2017glv,Billo:2018oog}. 
This is why the expansion (\ref{4.3}) 
is dubbed normal ordering and denoted by the usual double colon notation. 
%\begin{proof}
To prove this, we notice that 
by linearity we can reduce the problem 
to field monomial in the higher order traces $\tr(a^{n})$. Let $W(\Omega)$ be 
the operator constructed by subtraction of Wick pairings. The correlator $\langle W(\Omega)\,\Omega'\rangle$
is obtained after full Wick pairing. If $\dim \Omega'<\dim\Omega$, some pairing is necessarily inside 
$W(\Omega)$ and we get zero. By uniqueness, $\norm{\Omega}\equiv W(\Omega)$.
%\end{proof}

\paragraph{The case of  $\Omega_{n}$} 
We now consider the special case of the operators $\Omega_n = (\tr a^2)^n$. Their  expectation value is 
\be
\la{4.4}
\langle \Omega_n\rangle = \frac{\Gamma(\alpha+n)}{\Gamma(\alpha)},\qquad
\alpha=\frac{N^{2}-1}{2},
\ee
as follows easily from (\ref{3.3}). According the the remarks in the previous paragraph
% normal ordering amounts
%to subtraction of all Wick contractions. This implies that 
we can write the following specialized version  
of (\ref{4.3}) 
\be
\la{4.5}
\norm{\Omega_n} = \Omega_n + \sum_{k=0}^{n-1}\,c_k^{(n)}\,\Omega_k .
\ee
%where again the coefficients $c_k^{(n)}$ are determined by the conditions
%$\langle \norm{\Omega_n} \, \Omega_m\rangle = 0$, for $m=0, 1, \dots, n-1$.
We can give the explicit form of the coefficients $c_k^{(n)}$ as well as the inverse change of 
basis from $\norm{\bm\Omega^{(n)}}=\{\norm{\Omega_{k}}\}_{0\le k\le n}$ 
to $\bm\Omega^{(n)}=\{\Omega_{k}\}_{0\le k\le n}$. Indeed, we now show that
%\begin{framed}
\be
\la{4.6}
\norm{\Omega_n} 
= \sum_{k=0}^{n}\, (-1)^{n+k}\frac{\Gamma(\alpha+n)}{\Gamma(\alpha+k)}\,\binom{n}{k}
\,\Omega_k,\qquad 
\Omega_n 
= \sum_{k=0}^{n}\, \frac{\Gamma(\alpha+n)}{\Gamma(\alpha+k)}\,\binom{n}{k}
\,\norm{\Omega_k},
\ee
%\end{framed}
\noindent
where $\alpha$ has been defined in (\ref{4.4}). In the following it will be sometimes 
convenient to adopt an explicit matrix vector notation for 
(\ref{4.6}) and write
\be
\la{4.7}
\norm{\bm\Omega} = M\,\bm\Omega,\qquad M_{n,k} = c^{(n)}_{k}=
(-1)^{n+k}\frac{\Gamma(\alpha+n)}{\Gamma(\alpha+k)}\,\binom{n}{k}.
\ee
To prove (\ref{4.6}), we start
from the orthogonality condition 
$\langle \norm{\Omega_n} \, \Omega_m\rangle = 0$, with $m=0, 1, \dots, n-1$, that we write as 
\be
\langle\Omega_{n+m} + \sum_{k=0}^{n-1}\,c_k^{(n)}\,\Omega_{k+m}\rangle=0.
\ee
Using (\ref{4.4}), these conditions may be written
\be
\la{4.9}
\Gamma(\alpha+n+m) +  \sum_{k=0}^{n-1}\,c_k^{(n)}\,\Gamma(\alpha+k+m) = 0,\qquad 0\le m\le n-1.
\ee
The solution is unique and reads \footnote{
We can replace (\ref{4.10})  into (\ref{4.9}).
This gives the quantity $\Gamma(\alpha+m)\,\prod_{k=0}^{n-1}(m-k)$ that vanishes when 
the integer $m$ is such that $0\le m\le n-1$.}
\be
\la{4.10}
c_k^{(n)} = (-1)^{n+k}\frac{\Gamma(\alpha+n)}{\Gamma(\alpha+k)}\,\binom{n}{k}.
\ee
Notice that since $c_n^{(n)}=1$, we can write (\ref{4.5})  in the simpler form (\ref{4.6}). Finally,
we can prove that 
%
%%\be
%%\norm{\Omega_n} 
%%= \sum_{k=0}^{n}\, (-1)^{n+k}\frac{\Gamma(\alpha+n)}{\Gamma(\alpha+k)}\,\binom{n}{k}
%%\,\Omega_k.
%%\ee
%It is convenient to look at (\ref{4.6}) as a change of basis associated with the matrix $M$
%relating $\norm{\bm \Omega} = \{\norm{\Omega_{k}}\}$ to $\bm\Omega= \{\Omega_{k}\}$
%\be
%\la{x1}
%\norm{\bm\Omega} = M\,\bm\Omega,\qquad M_{n,k} = c^{(n)}_{k}=
%(-1)^{n+k}\frac{\Gamma(\alpha+n)}{\Gamma(\alpha+k)}\,\binom{n}{k}.
%\ee
%The inverse matrix is simply
\be
(M^{-1})_{n,k} = (-1)^{n+k}\,M_{n,k} = \frac{\Gamma(\alpha+n)}{\Gamma(\alpha+k)}\,\binom{n}{k},
\ee
by exploiting the algebraic relation
\be
\sum_{k=0}^n M_{n,k}(M^{-1})_{k,m} = \sum_{k=0}^n\,(-1)^{n+k}\frac{\Gamma(\alpha+n)}{\Gamma(\alpha+m)}\binom{n}{k}\binom{k}{m} = \frac{(-1)^{n+1}\sin(\pi m)}{\pi(n-m)}
\frac{\Gamma(\alpha+n)}{\Gamma(\alpha+m)},
\ee
and noting that the r.h.s. is zero for integer $m\neq n$ and one for $m\to n$. 

\paragraph{A useful elementary identity} 
The specific explicit coefficients (\ref{4.9}) allow to   prove the following elementary identity. For 
any function $F(g\,a)$ we can write
\be
\la{4.13}
\langle F\,\norm{\Omega_{n}}\rangle = g^{2n}\left(\frac{1}{2g}\frac{d}{dg}\right)^{n}\,\langle F\rangle.
\ee

\begin{proof} 
It is convenient to assume that $F(g\,a)$ may be expanded in a series of operators of the form (\ref{3.7}). 
\footnote{
This is already enough for our later use.  A more general proof is given in App.~\ref{app:gen}. }
Thus, we write for some set $\mc M$ of multi-indices
\be
\la{4.14}
F(g\,a) = \sum_{\bm m\in \mc M} c_{\bm m}\mc O_{\bm m}(g\,a) = 
\sum_{\bm m\in \mc M} c_{\bm m}\,g^{|\bm m|}\, \mc O_{\bm m}(a).
\ee
We compute $\langle \mc O_{\bm m}\,\Omega_{n}\rangle$ by iterating the  recursion 
relation proved in  \cite{Billo:2017glv}
\be
\la{4.15}
\langle \mc O_{\bm m}\,\Omega_{1}\rangle = \frac{N^{2}-1+|\bm m|}{2}\,\langle \mc O_{\bm m}\rangle.
\ee
This gives
\be
\langle \mc O_{\bm m}\,\Omega_{n}\rangle = \left(\prod_{k=1}^{n}\frac{N^{2}-1+|\bm m|+2(k-1)}{2}
\right)\,\langle \mc O_{\bm m}\rangle = \frac{\Gamma(\tfrac{|\bm m|}{2}+n+\alpha)}
{\Gamma(\tfrac{|\bm m|}{2}+\alpha)}\,\langle \mc O_{\bm m}\rangle.
\ee
Finally, using the mixing solution (\ref{4.6}), we have (for $|\bm m|>2n$ otherwise we get zero by construction
of $\norm{\Omega_{n}}$)
\begin{align}
\langle \mc O_{\bm m}\,\norm{\Omega_{n}}\rangle &= \sum_{k=0}^{n}(-1)^{n+k}\frac{\Gamma(\alpha+n)}
{\Gamma(\alpha+k)}\binom{n}{k}\frac{\Gamma(\tfrac{|\bm m|}{2}+k+\alpha)}
{\Gamma(\tfrac{|\bm m|}{2}+\alpha)}\,\langle \mc O_{\bm m}\rangle =
\frac{(\frac{1}{2}\,|\bm{m}|)!}{(\frac{1}{2}\,|\bm{m}|-n)!}\ \langle \mc O_{\bm m}\rangle,
\end{align}
Replacing this in  (\ref{4.14}) gives
\be
\langle F\,\norm{\Omega_{n}}\rangle = \sum_{\bm m\in \mc M} c_{\bm m}\,g^{|\bm m|}\,
\frac{(\frac{1}{2}\,|\bm{m}|)!}{(\frac{1}{2}\,|\bm{m}|-n)!}\ \langle \mc O_{\bm m}\rangle = 
g^{2n}\,\left(\frac{d}{d(g^{2})}\right)^{n}\langle F(g\,a)\rangle,
\ee
which is the same as (\ref{4.13}).
\end{proof}
 
 \paragraph{Remark 1} It may be interesting to compare  (\ref{4.13}) with 
 what is obtained by taking  derivatives with respect to $1/g^{2}$, {\em
i.e.} essentially with respect to $\text{Im}\tau$ where $\tau$ is the complexified gauge coupling.
In this case we would have  obtained Gaussian averages with various insertions of  powers of $\Omega_{1}$.
Instead, derivatives with respect to 
$g^{2}$ automatically build up the normal ordered operators $\norm{\Omega_{n}}$. In particular, 
the $N$ dependence of the normal ordering expansion (\ref{4.6})
is fully taken into account by the $N$ dependence of the basic expectation $\langle F\rangle$.

 \paragraph{Remark 2} 
 The relation  (\ref{4.13}) may be written in explicit form by simply rearranging derivatives. 
 This gives
 \be
\la{4.19}
\langle F\,\norm{\Omega_n}\rangle = \frac{(-1)^{n}}{2^{2\,n}}
\sum_{p=1}^{n}(-2)^{p}\,\frac{(2n-1-p)!}{(n-p)!(p-1)!}\ 
g^{p}\frac{d^{p}}{dg^{p}}\,\langle F\rangle.
\ee

\medskip\noindent
A useful  application of  formula (\ref{4.13}) is to the case when $F$ is the Wilson loop (\ref{3.8}) 
\be
W = \frac{1}{N}\sum_{k=0}^\infty\frac{g^k}{2^\frac{k}{2}\,k!}\,\tr a^k,
\ee
with expectation value  in the $SU(N)$ theory given by  \cite{Billo:2018oog} \footnote{
For the $U(N)$ theory, the center term $-g^{2}/(8N)$ in the exponent 
is absent \cite{Drukker:2000rr}.}
\be
\la{4.21}
\langle W \rangle = \frac{1}{N}\,\text{L}_{N-1}^1\left(-\frac{g^2}{4}\right)\,\exp\left[\frac{g^2}{8}
\left(1-\frac{1}{N}\right)\right].
\ee
Formula (\ref{4.13})  simply predicts
\be
\la{4.22}
\langle W\,\norm{\Omega_n}\rangle = g^{2n}\,\left(\frac{1}{2\,g}\frac{d}{dg}\right)^n\,\langle  W\rangle.
\ee
Special cases are conveniently obtained by applying (\ref{4.19}) 
\begin{align}
\langle W\,\norm{\Omega_{1}}\rangle &= \tfrac{g}{2}\,\partial_{g}\langle W\rangle,\notag \\
\langle W\,\norm{\Omega_{2}}\rangle &= \tfrac{g}{4}\,\partial_{g}\,(-1+g\,\partial_{g})\langle W\rangle,\notag \\
\langle W\,\norm{\Omega_{3}}\rangle &= \tfrac{g}{8}\,\partial_{g}\,(3-3\,g\,\partial_{g}+g^{2}\,\partial^{2}_{g}
)\,\langle W\rangle, \notag \\
\langle W\,\norm{\Omega_{4}}\rangle &= \tfrac{g}{16}\,\partial_{g}\,(
-15+15\,g\,\partial_{g}-6\,g^{2}\,\partial^{2}_{g}+g^{3}\,\partial^{3}_{g}
)\,\langle W\rangle, 
\end{align}
and so on. As a check, the first two lines are 
in agreement with Eqs.~(4.7, 4.11) of \cite{Billo:2018oog}.

Thanks to the special structure of $\langle W\rangle$ in  (\ref{4.21}), it is also simple to 
determine the closed expression of
 (\ref{4.22}) for  any given $N$, a kind of formula that will be useful in the following. The first cases are 
\begin{align}
\la{4.24}
\left. \langle W\,\norm{\Omega_{n}}\rangle\right|_{SU(2)} &= \left(\tfrac{g^{2}}{16}\right)^{n}\,
e^{\frac{g^{2}}{16}}\,(1+2n+\tfrac{1}{8}\,g^{2}), \notag \\
\left. \langle W\,\norm{\Omega_{n}}\rangle\right|_{SU(3)} &= \left(\tfrac{g^{2}}{12}\right)^{n}\,
e^{\frac{g^{2}}{12}}\,\big[1+\tfrac{3}{2}\,n+\tfrac{3}{2}\,n^{2}+\tfrac{1}{4}(1+n)\,g^{2}
+\tfrac{1}{96}\,g^{4}\big],\notag \\
\left. \langle W\,\norm{\Omega_{n}}\rangle\right|_{SU(4)} &=\left(\tfrac{3\,g^{2}}{32}\right)^{n}\,
e^{\frac{3\,g^{2}}{32}}
\big[
\tfrac{64 n^3+96 n^2+164 n+81}{81}+\tfrac{16 n^2+32 n+27}{72} g^2 +
\tfrac{2n+3}{96}\, g^4+\tfrac{g^6}{1536}\big].
\end{align}
The $SU(2)$ computation is  particularly simple and is easily checked by
direct evaluation of the partition function and correlators in the Gaussian matrix model. This is 
briefly reviewed in App.~\ref{app:su2}. The advantage of the
relations proved in this section is that they are parametric in $N$, {\em i.e.} in the gauge group rank.

\subsection{$\zeta(3)$ corrections in the $\N=2$ theory}

The correction factor arising in $\N=2$ correlators due to the first term in (\ref{3.4}) is
\be
1-\frac{3\,g^4}{64\,\pi^4}\,\zeta(3)\,\Omega_2+\dots = 1-\eps\,\Omega_2+\dots
\ee
where $\eps=\frac{3\,g^4}{64\,\pi^4}\,\zeta(3)$ is the term we want to control at linear order, and 
dots denote higher transcendentality structures. To 
compute its effects in correlators, we first need to determine how it affects normal ordering. To this aim,
we  need a simple  relation valid for the $\N=4$ normal ordered operators and expressing the 
decomposition of $\norm{\Omega_n}\,\Omega_2$ in terms of normal ordered operators
\begin{align}
\la{4.26}
\norm{\Omega_n}&\,\Omega_2 = 
\norm{\Omega_{n+2}}+2\,(\alpha+2n+1)\,\norm{\Omega_{n+1}}
+(\alpha^2+(6n+1) \alpha+6n^2)\,
\norm{\Omega_n}\notag \\
&+2\,n\,(\alpha+2n-1)(\alpha+n-1)\,\norm{\Omega_{n-1}}
+n(n-1)\,(\alpha+n-1)(\alpha+n-2)\,\norm{\Omega_{n-2}}.
\end{align}
This is obtained by using the special form of the matrix $M$ in (\ref{4.7})  and writing
\begin{align}
\norm{\Omega_n}\,\Omega_2 =  \sum_{m=0}^{n}M_{n,m}\Omega_m\Omega_2 = 
\sum_{m=0}^{n}\sum_{\ell=0}^{m+2}M_{n,m}(M^{-1})_{m+2,\ell}\,\norm{\Omega_\ell} \ .
\end{align} 
Replacing the explicit form of $M$ and $M^{-1}$ from (\ref{4.6}, \ref{4.7}) 
and doing the finite sums we get the above expansion (\ref{4.26}).
We  can now determine the $\eps$ perturbation
appearing in the $\N=2$ normal ordering
\begin{align}
\la{4.28}
\normtwo{\Omega_n} &= \sum_{k=0}^{n}\,(c_k^{(n)}+\eps\,d_k^{(n)}+\dots)\,\Omega_k,\notag \\
d_k^{(n)} &= (-1)^{k+n+1}\,(k+1)\,(2\alpha-1+3n+k)\,\frac{\Gamma(\alpha+n)}{\Gamma(\alpha+k)}\,
\binom{n}{k+1}.
\end{align}

\begin{proof}
Let us start with the obvious expansion \footnote{We remind that 
 $\llangle\cdots\rrangle$ denotes the expectation value in the $\N=2$ theory}
\begin{align}
\llangle A\rrangle &= \frac{\langle A\,(1-\eps\,\Omega_2+\dots)\rangle}
{\langle 1-\eps\,\Omega_2+\dots\rangle}= \langle A\rangle+\eps\,\bigg[
-\langle A\,\Omega_2\rangle + \langle A\rangle \langle \Omega_2\rangle 
\bigg]+\dots.
\end{align}
Now, let us consider $A = \normtwo{\Omega_{n}}X$ with $\dim X<2n$. If we write
$\normtwo{\Omega_{n}}=\norm{\Omega_{n}}+\delta\Omega_{n}$, we find
\be
\langle (\delta\Omega_{n}-\norm{\Omega_{n}}\Omega_{2})\,X\rangle = 0.
\ee
Due to the defining property of normal ordering, the explicit solution is 
\be
\la{4.31}
\delta\Omega_{n} = \norm{\Omega_{n}}\Omega_{2}-\norm{\Omega_{n+2}}-k\,\norm{\Omega_{n+1}}
-k'\,\norm{\Omega_{n}},
\ee
where the constant $k,k'$ are chosen such that $\delta\Omega_{n}$ does not contain $\Omega_{n+1}$
and $\Omega_{n}$. From (\ref{4.26}) we read
\be
k=2\,(2n+\alpha+1),\qquad k'= 6\,n^{2}+6\,n\,\alpha+\alpha\,(\alpha+1).
\ee
Rearranging (\ref{4.31}) proves formula (\ref{4.28}).
\end{proof}

\paragraph{Remark} Notice that in the r.h.s. of (\ref{4.28}) we find only operators of the form $\Omega_{n}$. 
This is  a special property of the $\zeta(3)$ correction and is false 
for corrections associated with higher transcendentality  terms.

\medskip
\noindent
Finally, we can compute the $\zeta(3)\sim \eps$ correction to the 
1-point function of the chiral primary $\normtwo{\Omega_{n}}$ with the $\frac{1}{2}$-BPS Wilson 
loop. The result is 
\be
\llangle W\,\normtwo{\Omega_n}\rrangle = \langle W\,\norm{\Omega_n}\rangle+\eps X_n+
\dots, 
\ee
with 
\begin{align}
\la{4.34}
X_n &= -f^{(n+2)}-2\,(\alpha+2n+1)\,f^{(n+1)}-6\,n\,(\alpha+n)\,f^{(n)},\notag \\
f^{(n)} &= g^{2n}\,\left(\frac{1}{2\,g}\frac{d}{dg}\right)^n\,\langle  W\rangle.
\end{align}

\begin{proof}
We have
\begin{align}
\la{4.35}
\llangle W\,\normtwo{\Omega_n}\rrangle &= 
\frac{\langle W\,\normtwo{\Omega_n}\,(1-\eps\,\Omega_2+\dots)\rangle}
{\langle 1-\eps\,\Omega_2+\dots\rangle}  = \langle W\,\normtwo{\Omega_n}\rangle+\eps\,\bigg[
-\langle W\,\normtwo{\Omega_n}\,\Omega_2\rangle + \langle W\,\normtwo{\Omega_n}\rangle \langle \Omega_2\rangle 
\bigg]+\dots\notag \\
&= \langle W\,\norm{\Omega_n}\rangle+\eps\,\bigg[
\sum_{k=0}^{n-1} d_k^{(n)}\langle W \Omega_k\rangle
-\langle W\,\norm{\Omega_n}\,\Omega_2\rangle + \langle W\,\norm{\Omega_n}\rangle \langle \Omega_2\rangle 
\bigg]+\dots.
\end{align}
Now we can use $\langle\Omega_2\rangle = \alpha\,(\alpha+1)$, and the decomposition (\ref{4.26}).
%
%\ee
%and
%\begin{align}
%\norm{\Omega_n}\,\Omega_2 =  \sum_{m=0}^{n}M_{n,m}\Omega_m\Omega_2 = 
%\sum_{m=0}^{n}\sum_{\ell=0}^{m+2}M_{n,m}(M^{-1})_{m+2,\ell}\,\norm{\Omega_\ell} \ .
%\end{align} 
%Using the explicit form of the matrix $M$, one obtains the important relation
%\begin{align}
%\norm{\Omega_n}&\,\Omega_2 = 
%\norm{\Omega_{n+2}}+2\,(\alpha+2n+1)\,\norm{\Omega_{n+1}}
%+(\alpha^2+(6n+1) \alpha+6n^2)\,
%\norm{\Omega_n}\notag \\
%&+2\,n\,(\alpha+2n-1)(\alpha+n-1)\,\norm{\Omega_{n-1}}
%+n(n-1)\,(\alpha+n-1)(\alpha+n-2)\,\norm{\Omega_{n-2}}.
%\end{align}
Finally, the first term in the square bracket in (\ref{4.35}) is manipulated using
\begin{align}
& \sum_{k=0}^{n-1}d_k^{(n)}\Omega_k = \sum_{k=0}^{n-1}d_k^{(n)}
\sum_{m=0}^{k}(M_{k,m})^{-1}\norm{\Omega_m}\notag \\
&\ \ = 2\,n\,(\alpha+2n)(\alpha+n-1)\,\norm{\Omega_{n-1}}
+n(n-1)(\alpha+n-1)(\alpha+n-2)\,\norm{\Omega_{n-2}}.
\end{align}
Using (\ref{4.22}) proves (\ref{4.34}).
\end{proof}

\noindent
Expanding at weak coupling, the leading term in the $X_{n}$ correction reads 
\begin{align}
X_{1} &= -\frac{3 (N^4-1)}{8 N}\,g^{2}+\dots, \notag\\
X_{2} &= -\frac{(N^2-1) (N^2+3) (2 N^2-3)}{32 N^2}\,g^{4}+\dots,\notag\\
X_{3} &= -\frac{3 (N^2-1) (N^2+5) (N^4-3 N^2+3)}{512 N^3}\,g^{6}+\dots,\notag\\
X_{4} &= -\frac{(N^2-1) (N^2+7) (2 N^6-8 N^4+15 N^2-15)}{5120 N^4}\,g^{8}+\dots,
\end{align}
and so on. One can check that, after multiplication by $\frac{3\zeta(3)}{(8\pi^{2})^{2}}\,g^{4}$,
the first two lines  give the entries labeled (2) and (2,2) 
in Tab.~(2) in \cite{Billo:2018oog}. 

\subsubsection{Checks in the planar limit}

From the result (\ref{4.34})
it is easy to work out the planar limit $N\to \infty$ with fixed $\lambda = N\,g^2$. Indeed we can expand
\be
f^{(n)} = \lambda^n\,\frac{d^n}{d\lambda^n}\,\langle W\rangle = f^{(n)}_0+\mc O(1/N^2),\qquad
f_0^{(n)} = \lambda^n\,\frac{d^n}{d\lambda^n}\bigg[\frac{2}{\sqrt\lambda}\,\text{I}_1(\sqrt\lambda)\bigg].
\ee
In (\ref{4.34}), we have at leading order 
\be
X_n = -N^2\,(f_0^{(n+1)}+3\,n\,f_0^{(n)})+\mc O(N^0),
\ee
and therefore
\be
\la{4.40}
\llangle W\normtwo{\Omega_n}\rrangle = \bigg\{\lambda^n\,\frac{d^n}{d\lambda^n}
-\frac{3\,g^4\zeta(3)}{64\pi^4}N^2\,\bigg[\lambda^{n+1}\,\frac{d^{n+1}}{d\lambda^{n+1}}
+3\,n\,\lambda^{n}\,\frac{d^{n}}{d\lambda^{n}}\bigg]+\dots\bigg\}\frac{2}{\sqrt\lambda}\,\text{I}_1(\sqrt\lambda).
\ee
Rearranging derivatives, we can write (\ref{4.40})  in the form 
\be
\la{4.41}
\llangle W\normtwo{\Omega_n}\rrangle = \bigg\{1
-\frac{3\,\zeta(3)}{64\pi^4}\,\lambda^2\,\bigg[\lambda\,\frac{d}{d\lambda}
+2\,n\bigg]+\dots\bigg\}
\lambda^n\,\frac{d^n}{d\lambda^n}\bigg[\frac{2}{\sqrt\lambda}\,\text{I}_1(\sqrt\lambda)\bigg].
\ee
Sample cases are 
\begin{align}
& \llangle W\rrangle = \frac{2}{\sqrt\lambda}\,\text{I}_1(\sqrt\lambda)-
\frac{3\,\zeta(3)}{(8\pi^2)^2}\,\lambda^2\,\text{I}_2(\sqrt\lambda)+\dots, \notag \\
& \llangle W\normtwo{\Omega_1}\rrangle = \text{I}_2(\sqrt\lambda)-
\frac{3\,\zeta(3)}{(8\pi^2)^2}\,\bigg[
\lambda^2\,\text{I}_0(\sqrt\lambda)+\tfrac{1}{2}\lambda^{3/2}(\lambda-4)\,\text{I}_1(\sqrt\lambda)
\bigg], \notag \\
& \llangle W\normtwo{\Omega_2}\rrangle = -2\text{I}_0(\sqrt\lambda)
+\frac{8+\lambda}{2\sqrt\lambda}\,\text{I}_1(\sqrt\lambda)
-\frac{3\,\zeta(3)}{(8\pi^2)^2}\,\bigg[
\tfrac{\lambda-24}{4}\,\lambda^2\,\text{I}_0(\sqrt\lambda)
+\lambda^{3/2}(\lambda+12)\,\text{I}_1(\sqrt\lambda)
\bigg]+\dots.
\end{align}
These may be compared with the general formula conjectured in \cite{Billo:2018oog} that reads
\begin{align}
\llangle W\normtwo{\Omega_n}\rrangle &= \frac{\lambda^{\frac{n-1}{2}}}{2^{n-1}}\,
\text{I}_{n+1}(\sqrt\lambda)-\frac{3\,\zeta(3)}{(8\pi^2)^2}\,\frac{\lambda^\frac{n+4}{2}}{2^n}
\bigg[\text{I}_n(\sqrt\lambda)+\frac{2\,(2n-1)}{\sqrt\lambda}\,\text{I}_{n+1}(\sqrt\lambda)\bigg]+\dots.
\end{align}
and, of course, there is agreement after some rearrangement of the Bessel functions.

\section{Large R-charge limit and universality}
\la{sec:large}

We now make full use of the  results presented in the  previous section to discuss the large R-charge limit of the 
normalized one-point function of the chiral primary $\Omega_{n}(\varphi)$ with the $\tfrac{1}{2}$-BPS
Wilson loop in the
$\N=2$ theory. In particular, the result  (\ref{4.34})  controls exactly the contribution $\sim \zeta(3)$.
When combined with explicit low $N$ calculations, 
it  provides a guide to understand the behaviour of higher  transcendentality contributions, as we are 
going to illustrate.

\subsection{Analysis of the $\zeta(3)$ correction}

We define the asymptotic ratio of one-point functions, cf. (\ref{4.2}),
\be
\la{5.1}
G(\kappa; N) = 
\lim_{n\to \infty}\left. \frac{\llangle W\,\normtwo{\Omega_{n}}\rrangle}{\langle W\,\norm{\Omega_{n}}\rangle}
\right|_{\kappa=ng^{2}}.
\ee
The very existence of the large $n$ limit in (\ref{5.1}) 
is something that deserves investigation. 
To begin our analysis, let 
us consider the simplest $SU(2)$ case. Using  (\ref{4.41})  and the first equation in (\ref{4.24}) we obtain the  
ratio
\begin{align}
\la{5.2}
\frac{\llangle W\,\normtwo{\Omega_{n}}\rrangle}{\langle W\,\norm{\Omega_{n}}\rangle} &= 
1-\frac{3 g^4}{16384\pi^{4}\,(8+16n+g^{2})}
\notag \\
&\bigg[6144 n (2 n+1) (2 n+3)+640 g^2 (2 n+1) (2 n+3)+40 g^4 (2 n+3)+g^6
\bigg]\,\zeta(3)+\text{h.z.},
\end{align}
where ``h.z.'' stands for higher transcendentality terms $\zeta(5)$, $\zeta(3)^{2}$, and so on.
The expression (\ref{5.2}) is exact in the gauge coupling $g$, {\em i.e.} it 
resums all contributions $\sim\zeta(3)\,g^{4+2\,k}$, with $k=0,1,2,\dots$.
For large $n$ and fixed $\kappa=n\,g^{2}$ the limit (\ref{5.1})  reads
\be
\la{5.3}
G(\kappa; 2)= 1-\frac{9\zeta(3)}{32\pi^{4}}\,\kappa^{2}+\mc O(\kappa^{3}),
\ee
and comes entirely from the $\zeta(3)\,g^{4}$ term in (\ref{5.2}).
The same analysis may be repeated for higher  $N$, see (\ref{4.24}). In all cases one finds the same
leading behaviour as in (\ref{5.3}). In general, one can show that for 
$SU(N)$ the $\zeta(3)$ correction reads
\begin{align}
\la{5.4}
\frac{\llangle W\,\normtwo{\Omega_{n}}\rrangle}{\langle W\,\norm{\Omega_{n}}\rangle} &= 
1-\frac{9\,n\,(2\,n+N^{2}-1)\,\zeta(3)}{64\pi^{4}}\,\,\bigg[g^{4}+\mc O(g^{6})\bigg]+\text{h.z.},
\end{align}
and thus (\ref{5.3}) is valid for any $N$. Indeed, our results allow to check that the higher order
terms in coupling $g$ are always accompanied by lower powers of $n$ and do not contribute in the 
$n\to \infty$ limit. Notice however, that the complexity of the detailed dependence on $n$ increases with $N$.
For instance, even for the simplest $\zeta(3)$ correction, the $\mc O(g^{6})$ term in (\ref{5.4})
reads for $SU(N)$ with $N=2, 3, 4$
{\small
\begin{align}
\la{5.5}
\left. \frac{\llangle W\,\normtwo{\Omega_{n}}\rrangle}{\langle W\,\norm{\Omega_{n}}\rangle}
\right|_{SU(2)} &= 
1-\frac{9\,n\,(2\,n+3)\,\zeta(3)\,g^{4}}{64\pi^{4}}\,\bigg[
1+\frac{g^2 (4 n+5)}{48 n (2 n+1)}+\mc O(g^{4})\bigg]+\text{h.z.},\notag \\
\left. \frac{\llangle W\,\normtwo{\Omega_{n}}\rrangle}{\langle W\,\norm{\Omega_{n}}\rangle}
\right|_{SU(3)} &= 1-\frac{9\,n\,(2\,n+8)\,\zeta(3)\,g^{4}}{64\pi^{4}}\,\bigg[
1+\frac{g^2 (2 n+5) (3 n^2+9 n+8)}{36 n (n+4) (3 n^2+3 n+2)}+\mc O(g^{4})
\bigg]+\text{h.z.}, \\
\left. \frac{\llangle W\,\normtwo{\Omega_{n}}\rrangle}{\langle W\,\norm{\Omega_{n}}\rangle}
\right|_{SU(4)} &= 1-\frac{9\,n\,(2\,n+15)\,\zeta(3)\,g^{4}}{64\pi^{4}}\,\bigg[
1+\frac{g^2 (4 n+17) (64 n^3+288 n^2+548 n+405)}{32 n (2 n+15) (64 \
n^3+96 n^2+164 n+81)}+\mc O(g^{4})
\bigg]+\text{h.z.}.\notag
\end{align}
}

\noindent
We cannot give a general formula parametric in $N$, but in 
all cases, the correction term in the square brackets goes like $g^{2}/n = \kappa/n^{2}$
and is negligible at large $n$.

\subsection{Higher transcendentality terms}

The agreement of (\ref{5.3}) with the analogous universal $\zeta(3)$ term in the 
chiral 2-point scaling function $F(\kappa; N)$, 
cf. (\ref{2.7})  and (\ref{2.8}), is a strong motivation to look at higher transcendentality terms.
%For them, we do not have closed expressions as for the simpler $\zeta(3)$ correction. Nevertheless, 
We can 
work out the first values of the ratio $\llangle W\,\normtwo{\Omega_{n}}\rrangle/
\langle W\,\norm{\Omega_{n}}\rangle$ looking at the first appearance of a certain transcendentality term 
$\zeta(\bm{s})$, cf. (\ref{2.5}), working at fixed small $N$. In other words, we shall 
assume that the $\zeta(3)$ pattern repeats for higher transcendentality structures. In particular, for a given
structure,  higher powers of $g$ will be associated to subleading terms for $n\to \infty$, as in (\ref{5.5}). 
\footnote{
These assumptions
may be examined in details for the $SU(2)$ and $SU(3)$ theories, 
where it is straightforward to compute the explicit integral definition of the partition function, as we
did for the $\N=4$ $SU(2)$ theory in App. \ref{app:su2}. }

\paragraph{SU(2) theory}
Working out the 
mixing algorithm for $\N=2$ correlators and evaluating
the correction to the 1-point function with the Wilson loop with $n=1,\dots, 20$, we find the following 
$\zeta(5)$ correction in $SU(2)$ case (extending (\ref{5.2}) that was for the $\zeta(3)$ term)
\begin{align}
\la{5.6}
& \left. \frac{\llangle W\,\normtwo{\Omega_{n}}\rrangle}{\langle W\,\norm{\Omega_{n}}\rangle} 
\right|_{SU(2), \zeta(5)\ \text{term}} = \frac{5\,g^{6}\,\zeta(5)}{2097152\,(8+16n+g^{2})\,\pi^{6}}
\bigg[163840\ n (2 n+1) (4 n^2+9 n+8)\notag \\
&\quad +17920 g^2 (2 n+1) (4 n^2+10 n+9)+2688 \
g^4 (2 n^2+6 n+5)+112 g^6 (n+2)+g^8\bigg].
\end{align}
As in (\ref{5.3}), the lowest power of $g$ is the only one surviving the $n\to \infty$ limit at 
fixed $n\,g^{2}$. Extending the analysis to higher transcendentality structures and writing only the 
terms that give contribution in the $n\to \infty$ limit, we obtain, 
\begin{align}
\la{5.7}
%& \text{\bf SU(2)}\notag \\
& \left. \frac{\llangle W\,\normtwo{\Omega_{n}}\rrangle}{\langle W\,\norm{\Omega_{n}}\rangle}
\right|_{SU(2)} =
1-9\,n\, (2 n+3)\,\zeta(3)\,\left[\ghat^{2}+\dots\right]
+25 \, n\, (4 n^2+9 n+8)\,\zeta(5)\,\left[\ghat^{3}+\dots\right]\notag \\
&+\bigg[\tfrac{27}{2} n (12 n^3+60 n^2+81 n+47)\,\zeta(3)^{2}
-\tfrac{2205}{16}  n (2 n+3) (2 n^2+3 n+4)\,\zeta(7)\bigg]\,\left[\ghat^{4}+\dots\right]\notag \\
&+\bigg[-\tfrac{225}{2}  n (2 n+3) (8 n^3+42 n^2+52 n+45)\,\zeta(3)\zeta(5)\notag \\
&+\tfrac{3213}{32}  n (32 n^4+120 n^3+240 n^2+270 n+163)\,\zeta(9)
\bigg]\,\left[\ghat^{5}+\dots\right]\notag \\
&+\bigg[
-\tfrac{243}{2}  n (2 n+3) (4 n^4+36 n^3+103 n^2+107 n+72)\,\zeta(3)^{3}\notag \\
&+\tfrac{125}{16}  n (640 n^5+5760 n^4+16600 n^3+26400 \
n^2+24700 n+12777)\,\zeta(5)^{2}\notag \\
&+\tfrac{19845}{16} n (8 n^5+68 n^4+190 n^3+307 n^2+288 n+147)\,\zeta(3)\,\zeta(7)\notag \\
&-\tfrac{78771}{256}  n (2 n+3) (4 n^2+6 n+11) (8 n^2+12 n+19)\,\zeta(11)
\bigg]\,\left[\ghat^{6}+\dots\right]+\text{h.z.}\ ,
\end{align}
where dots stand for higher order \underline{rational} corrections in $g$. In other words, (\ref{5.7}) 
gives the exact $n$ dependence of the lowest order correction (in $g$) to the structures 
\be
\zeta(3), \ \zeta(5), \ \zeta(3)^{2},\ \zeta(7), \ \zeta(3)\zeta(5),\ \zeta(3)^{3},\ \zeta(5)^{2},\ \zeta(3)\zeta(7),\
\zeta(11).
\ee
Taking  the $n\to\infty$ limit, one obtains 
\begin{align}
G(\kappa; 2) &= 
\lim_{n\to \infty}\left. \frac{\llangle W\,\normtwo{\Omega_{n}}\rrangle}{\langle W\,\norm{\Omega_{n}}\rangle}
\right|_{\kappa=ng^{2}}= 1-\frac{9  \zeta (3)}{32 \pi ^4}\kappa ^2+\frac{25 \zeta 
(5)}{128 \pi ^6} \kappa ^3+ \bigg(\frac{81 \zeta (3)^2}{2048 \pi \
^8}-\frac{2205 \zeta (7)}{16384 \pi ^8}\bigg)\kappa ^4\notag \\
&+\bigg(\frac{3213 \zeta 
(9)}{32768 \pi ^{10}}-\frac{225 \zeta (3) \zeta (5)}{4096 \pi 
^{10}}\bigg)\,\kappa ^5 +\bigg (-\frac{243 \zeta (3)^3}{65536 \pi ^{12}}+\frac{625 
\zeta (5)^2}{32768 \pi ^{12}}\notag \\
&+\frac{19845 \zeta (3) \zeta (7)}{524288 
\pi ^{12}}-\frac{78771 \zeta (11)}{1048576 \pi ^{12}}\bigg)\,\kappa^{6}+\mc O(\kappa ^7),
\end{align}
and this nicely exponentiates to 
\be
\la{5.10}
\log G(\kappa; 2) = -\frac{9\, \zeta (3)}{32 \pi ^4}\, \kappa ^2 +\frac{25\,  \zeta 
(5)}{128 \pi ^6}\,\kappa ^3-\frac{2205\,  \zeta (7)}{16384 \pi 
^8}\,\kappa ^4+\frac{3213\,  \zeta (9)}{32768 \pi ^{10}}\,\kappa ^5
 -\frac{78771 \,
 \zeta (11)}{1048576 \pi ^{12}}\,\kappa ^6+\mc O(\kappa^{7}),
\ee
with only simple zeta functions. Comparing with (\ref{2.7}) we see that $G(\kappa; 2)$ coincides
with $F(\kappa; 2)$ at this order. We are thus led to the 
following 
\be
\la{5.11}
\text{conjecture:} \quad G(\kappa; 2) = F(\kappa; 2) \equiv \mc F(\kappa),
\ee
that our calculation supports  at order $\kappa^{6}$ included. The importance of (\ref{5.11}) 
as a guiding remark will be manifest in the next section.
Before elaborating on this, it is interesting to explore what happens for 
higher rank gauge groups. 

\paragraph{SU(3) theory}
Already the study of the first case $N=3$ reveals some interesting feature of the $G$ function.
We may check that again, higher transcendentality terms are dominated by the lowest order contribution in $g$
as soon as $n\to \infty$. For instance, the equation corresponding to (\ref{5.6}) reads in $SU(3)$ case
\begin{align}
\la{5.12}
 &\left. \frac{\llangle W\,\normtwo{\Omega_{n}}\rrangle}{\langle W\,\norm{\Omega_{n}}\rangle} 
\right|_{SU(3), \zeta(5)\ \text{term}} = \frac{5\,g^{6}\,\zeta(5)}{7962624\,
[48 (3 n^2+3 n+2)+24 g^2 (n+1)+g^4]\,\pi^{6}}\notag \\
& \qquad
\bigg[
829440 n (42 n^4+375 n^3+850 n^2+927 n+254)\notag \\
&\qquad +103680 g^2 (77 n^4+695 \
n^3+1746 n^2+1562 n+408)\notag \\
&\qquad+1728 g^4 (392 n^3+3213 n^2+6971 n+4080)+48 \
g^6 (588 n^2+3801 n+5288)\notag\\
&\qquad +12 g^8 (56 n+209)+7 g^{10}
\bigg].
\end{align}
Keeping only the relevant terms in large $n$ limit and working out higher transcendentality structures up
to $\zeta(11)$ the $SU(3)$ expansion corresponding to (\ref{5.7}) has  a similar structure
{\small
\begin{align}
%& \text{\bf SU(3)}\notag \\
& \left. \frac{\llangle W\,\normtwo{\Omega_{n}}\rrangle}{\langle W\,\norm{\Omega_{n}}\rangle}
\right|_{SU(3)} =
1-18 n (n+4)\zeta(3)\,\left[\ghat^{2}+\dots\right]\notag \\
&+\frac{50 n (42 n^4+375 n^3+850 n^2+927 n+254)}{9 (3 n^2+3 n+2)}\zeta(5)\,\left[\ghat^{3}+\dots\right]\notag \\
&+\bigg[18 n (9 n^3+90 n^2+252 n+199)\zeta(3)^{2}\notag \\
&
-\frac{245 n (147 n^5+1971 n^4+6815 n^3+12817 n^2+11122 n+2168)}{36 \
(3 n^2+3 n+2)}\zeta(7)\bigg]\,\left[\ghat^{4}+\dots\right]\notag \\
&+\bigg[-\frac{100 n (42 n^6+669 n^5+3754 n^4+9073 n^3+12914 n^2+7772 \
n+2496)}{3 n^2+3 n+2}\zeta(3)\zeta(5)\notag \\
&+\frac{7 n (3834 n^6+72063 n^5+354855 n^4+885855 n^3+1388231 \
n^2+916002 n+195160)}{6 (3 n^2+3 n+2)}\zeta(9)
\bigg]\,\left[\ghat^{5}+\dots\right]\notag \\
&+\bigg[-108 n (n+4) (9 n^4+126 n^3+558 n^2+957 n+632)
\zeta(3)^{3}\notag \\
&+\frac{250 n (2940 n^7+62790 n^6+518341 n^5+1817865 n^4+3833881 \
n^3+4405191 n^2+2647324 n+758148)}{81 (3 n^2+3 n+2)}\zeta(5)^{2}\notag \\
&+\frac{245 n (441 n^7+9441 n^6+71925 n^5+250551 n^4+511026 n^3+625772 \
n^2+342468 n+101560)}{6 (3 n^2+3 n+2)}\zeta(3)\,\zeta(7)\notag \\
&-\frac{77}{324 (3 n^2+3 \
n+2)}\, n (87549 n^7+2205555 n^6+14721010 n^5+47885868 \
n^4+101153611 n^3\notag \\
&+122829225 n^2+76171270 n+16090872)\zeta(11)
\bigg]\,\left[\ghat^{6}+\dots\right]+\text{h.z.}\ .
\end{align}
}
In the $n\to \infty$ limit we find 
{\small
\begin{align}
G(\kappa; 3) &=  1-\frac{9  \,\zeta (3)}{32 \,\pi ^4}\kappa ^2
+\frac{175  \,\zeta 
(5)}{1152 \,\pi ^6}\kappa ^3+\bigg (\frac{81 \,\zeta (3)^2}{2048 \,\pi 
^8}-\frac{12005 \,\zeta (7)}{147456 \,\pi ^8}\bigg)\,\kappa^{4}+ \bigg(\frac{1491 
\,\zeta (9)}{32768 \,\pi ^{10}}-\frac{175 \,\zeta (3) \,\zeta (5)}{4096 \,\pi 
^{10}}\bigg)\,\kappa^{5}\notag \\
&+\bigg (-\frac{243 \,\zeta (3)^3}{65536 \,\pi 
^{12}}+\frac{30625 \,\zeta (5)^2}{2654208 \,\pi ^{12}}+\frac{12005 \,\zeta 
(3) \,\zeta (7)}{524288 \,\pi ^{12}}-\frac{2247091 \,\zeta (11)}{84934656 
\,\pi ^{12}}\bigg)\,\kappa^{6}+\mc O(\kappa ^7).
\end{align}
}
Taking the logarithm we obtain again a neat exponentiation involving only terms linear in $\zeta$ functions
\be
\log G(\kappa; 3) = 
-\frac{9 \,\zeta (3)}{32 \,\pi ^4}\kappa ^2 +\frac{175 \,\zeta 
(5)}{1152 \,\pi ^6} \kappa ^3-\frac{12005\,\zeta (7)}{147456 \,\pi 
^8} \kappa ^4 +\frac{1491  \,\zeta (9)}{32768 \,\pi ^{10}}\kappa ^5
-\frac{2247091  \,\zeta (11)}{84934656 \,\pi ^{12}}\kappa ^6+\mc O(\kappa ^7).
\ee
Comparing with (\ref{2.8}), we see that $G(\kappa; 3) \neq F(\kappa; 3)$. In particular, the scaling 
function $G$ is simpler since exponentiates as in $N=2$ case.

\section{The all-order $SU(2)$ large R-charge scaling function: A conjecture}
\la{sec:all}

Given the striking universal presence of the scaling function $F(\kappa; 2)\equiv \mc F(\kappa)$, we now
present a
puzzling conjecture about its all-order expansion.
In (\ref{2.7}) and (\ref{5.10}) we already presented the expansion of its logarithm. 
For the reader's advantage, let us repeat here the first four terms
\begin{align}
\la{6.1}
\log & \mc F(\kappa) = -\frac{9\, \zeta (3)}{32 \,\pi ^4}\, \kappa ^2 +\frac{25\,  \zeta 
(5)}{128 \,\pi ^6}\,\kappa ^3-\frac{2205\,  \zeta (7)}{16384 \,\pi 
^8}\,\kappa ^4+\frac{3213\,  \zeta (9)}{32768 \,\pi ^{10}}\,\kappa ^5+\dots.
\end{align}
As discussed in \cite{Bourget:2018obm} 
the special feature of $SU(2)$ is the linear dependence on the $\zeta$ functions. \footnote{As we have
discussed, this is a special feature of the $F(\kappa; 2)$ function that is not valid for $N>2$. However,
it seems not to be a characterization
of $G(\kappa; N)$ for $N>2$.}
In the $\N=2$ matrix model there is another natural and universal object with this property, {\em i.e.} the 
interaction action $S_{\text{int}}$ in (\ref{3.4}). Evaluating the expectation value $\langle S_{\rm int}
\rangle$ in the Gaussian matrix model associated with the $SU(N)$ theory gives 
\begin{align}
\la{6.2}
\langle S_{\rm int}\rangle &= \frac{3 (N^4-1) \,\zeta (3)}{256 \,\pi ^4}g^4 -\frac{5  (2 N^2-1) 
(N^4-1) \,\zeta (5)}{2048 \,\pi ^6 N}g^6+\frac{35  (N^2-1) (8 N^6+4 
N^4-3 N^2+3) \,\zeta (7)}{131072 \,\pi ^8 N^2}g^8\notag \\
&-\frac{21  (N^2-1) 
(26 N^8+28 N^6-3 N^4+6 N^2-9) \,\zeta (9)}{524288 \,\pi ^{10} 
N^3}g^{10}\notag \\
&+\frac{77  (N^2-1) (122 N^{10}+280 N^8+48 N^6-15 N^4+45) 
\,\zeta (11)}{16777216 \,\pi ^{12} N^4}g^{12}+\mc O(g^{14}).
\end{align}
The $N=2$ specialization  of (\ref{6.2}) is
\begin{align}
\langle S_{\rm int}\rangle &= \frac{45 \,\zeta (3)}{256 \,\pi ^4}g^4 
-\frac{525  \,\zeta (5)}{4096 \,\pi ^6}g^6+\frac{59535  \,\zeta (7)}{524288 \,\pi ^8}g^8-\frac{530145  
\,\zeta (9)}{4194304 \,\pi ^{10}}g^{10}+\frac{46081035  \,\zeta 
(11)}{268435456 \,\pi ^{12}}g^{12}+\mc O(g^{14}).
\end{align}
Comparing with (\ref{6.1}) we observe the following remarkably simple relation 
\be
\la{6.4}
\log \mc F(\kappa) = \langle S_\text{int}\rangle  \quad\text{with the replacement}\ \ 
g^{2n}\to -\frac{4^{n}}{(1+2n)\,n!}\,\kappa^{n},
\ee
where the replacement has to be done after expansion in powers of $g^{2}$. Assuming this relation to be 
correct at all orders we can evaluate $\langle S_{\rm int}\rangle$ using the formulas of App.~(\ref{app:su2}).
This gives
\begin{align}
\log \mc F(\kappa) &= \frac{4\sqrt 2}{\sqrt\pi}\int_{\mathbb R}da\ a^{2}\ e^{-2a^{2}}\ 
\sum_{n=2}^{\infty}(-1)^{n}\,\frac{2\,(4^{n}-4)}{n}\,\zeta(2n-1)
%\times \notag \\
%&\times 
\left(\frac{1}{8\pi^{2}}\right)^{n}\,a^{2n}\, \frac{-4^{n}}{(1+2n)\,n!}\,\kappa^{n}.
\end{align}
Doing the trivial integrals, we obtain a closed form for all perturbative coefficients of 
$\log \mc F(\kappa)$ 
%\begin{framed}
\be
\la{6.6}
\log \mc F(\kappa) = 4\,\sqrt\pi\,\sum_{n=2}^{\infty}(-1)^{n+1}\,\frac{4^{n}-4}{n\,n!\,(2\pi)^{2n+1}}\,
\Gamma(n+\tfrac{1}{2})\,\zeta(2n-1)\,\kappa^{n}.
\ee
%\end{framed}
One can easily check that the additional terms beyond those in (\ref{6.1}) fully reproduce the 
$\mc O(\kappa^{10})$ result (\ref{2.7}) giving strong support to the validity of (\ref{6.6}), {\em i.e.}
of the replacement rule (\ref{6.4}). 

\medskip
\noindent
The series (\ref{6.6}) has a finite convergence radius,
being convergent  for $|\kappa|<\pi^{2}$. A simple representation of its analytic continuation that may be used
for any $\kappa>0$ is 
\be
\la{6.7}
\log \mc F(\kappa) =\frac{1}{\pi^{2}} \int_{0}^{\infty}\frac{dt}{t^{2}}\ 
\frac{4\pi^{2}\,J_{0}(\frac{t\sqrt\kappa}{\pi})+4\pi\,t\,\sqrt\kappa\,J_{1}(\frac{t\sqrt\kappa}{\pi})
-\kappa\,t^{2}-4\pi^{2}}
{e^{t}+1},
\ee
where $J_{p}$ are standard Bessel functions. For $\kappa\gg 1$, one has  $\log \mc F(\kappa)  =
-\frac{\log 2}{\pi^{2}}\,\kappa+\mc O(\log\kappa)$, see App.~\ref{app:asy} for more information.
Of course, this large $\kappa$ expansion must be taken with some caution
because $\mc F(\kappa)$ does not include the
 instanton contribution. Instantons are expected not to contribute
at finite $\kappa$ -- and in particular in the weak coupling expansion -- 
but may be important when attempting to reach the large $\kappa$ regime.

\appendix
\section{Direct evaluation of $\N=4$ correlators in the $SU(2)$ theory}
\la{app:su2}

The analysis of the low gauge group rank cases may be carried on explicitly by direct evaluation of the 
partition function, at least in the simple Gaussian case, {\em i.e.} for the $\N=4$ theory. Let us briefly 
discuss the $SU(2)$ case as an illustration and to write down some expressions that are used in the main text.
The partition function (\ref{3.3}) may be written in terms of $a_{k}$, $k=1,\dots,N$, the eigenvalues of  $a$ 
\footnote{The normalization $\mc C$ depends on $g$ but drops in correlation functions.}
\begin{align}
\la{A.1}
Z_{S^{4}} &= \mc C\,\int\prod_{k=1}^{N}da_{k}\ \delta\big(\sum_{k} a_{k}\big)\,
\prod_{k<\ell}(a_{k}-a_{\ell})^{2} \ e^{-S_\text{int}}\ 
e^{-\sum a^{2}}, 
\end{align}
where
\begin{align}
\la{A.2}
S_{\text{int}}& = -\log\frac{\prod_{k<\ell=1}^{N}h(a_{k}-a_{\ell},g)^{2}}
{\prod_{k=1}^{N}h(a_{k}, g)^{2N}},\notag \\
\log h(x,g) &= (1+\gamma_\text{E})\,\frac{g^{2}}{8\pi^{2}}\,x^{2}+
\sum_{n=2}^{\infty}\frac{(-1)^{n+1}}{n}\zeta(2n-1)\left(\frac{g^{2}}{8\pi^{2}}\right)^{n}\,x^{2n}
\end{align}
In the special case $N=2$ we have  (with a suitable redefinition of $\mc C$),
\begin{align}
\la{A.3}
Z_{S^{4}} &= \mc C\,\int_{\mathbb R} da\ a^{2} \ e^{-S_\text{int}}\ 
e^{-2a^{2}}, \qquad
S_{\rm int} = \sum_{n=2}^{\infty}(-1)^{n}\,\frac{2\,(4^{n}-4)}{n}\zeta(2n-1)\left(\frac{g^{2}}{8\pi^{2}}\right)^{n}\,a^{2n}.
\end{align}
The Wilson loop operator is simply
\be
\la{A.4}
W(a) = \cosh\left(\frac{g}{\sqrt 2}\,a\right),
\ee
and the normal ordered operator $\Omega_{n}$ in $\N=4$ may be written, cf. (\ref{4.6}),
\be
\norm{\Omega_{n}} = (-1)^{n}\,n!\,\text{L}_{n}^{1/2}(2\,a^{2}) = \frac{1}{2^{2n+1}}
\frac{H_{2n+1}(a\,\sqrt{2})}{a\,\sqrt{2}},
\ee
where $\text{L}_{n}^{1/2}$ is the generalized Laguerre polynomial $\text{L}_{n}^{q}$ with $q=1/2$, and 
$H_{n}(z)$ are Hermite polynomials.
As a check one can compute
\begin{align}
\langle W \norm{\Omega_{n}}\rangle &= \frac{1}{2^{2n+1}}\,\frac{4\sqrt 2}{\sqrt\pi}\int_{\mathbb R}
da\,a^{2}\,\cosh\left(\frac{g}{\sqrt 2}\,a\right)\,\frac{H_{2n+1}(a\,\sqrt{2})}{a\,\sqrt{2}}\,
e^{-2a^{2}} \notag \\
&= \left(\tfrac{g^{2}}{16}\right)^{n}\,
e^{\frac{g^{2}}{16}}\,(1+2n+\tfrac{1}{8}\,g^{2}),
\end{align}
in agreement with (\ref{4.24}).

\section{General proof of Eq.~(\ref{4.13})}
\la{app:gen}

A general proof of the relation (\ref{4.13}) that does not assume $F$ to be represented in terms
of $\mc O_{\bm m}$ operators is as follows. For any $g$-independent function $F(a)$, let us denote by $\langle\cdot\rangle_{0}$ the 
Gaussian model expectation value after scaling $a\to a/g$
\be
\langle F(a) \rangle_0 = \frac{\int Da\, F(a)\, \exp\big(-\frac{1}{g^2}\,\Omega_1\big)}
{\int Da\,  \exp\big(-\frac{1}{g^2}\,\Omega_1\big)},
\ee
Clearly, we have
\be
\langle F(a)\ \Omega_n\rangle_0 = g^{2n}\langle F(g\,a)\ \Omega_n\rangle.
\ee
Now, let us assume that (\ref{4.13})  holds, with the expansion (\ref{4.7}) in the l.h.s.
We find 
\begin{align}
\langle & F \norm{\Omega_{n+1}}\rangle = g^{2(n+1)}\frac{d}{dg^{2}}\bigg[
g^{-2n}\langle F \norm{\Omega_{n}}\rangle\bigg] =
g^{2(n+1)}\frac{d}{dg^{2}}\bigg[
g^{-2n}\langle F \sum_{k=0}^{n}c_{k}^{(n)}\Omega_{k}\rangle\bigg] \notag \\
&= g^{2(n+1)}\sum_{k=0}^{n}c_{k}^{(n)} \frac{d}{dg^{2}}\bigg[g^{-2n-2k}
\langle F \Omega_{k}\rangle_{0}\bigg] \notag \\
&= -\sum_{k=0}^{n}c_{k}^{(n)}(n+k)\langle F \Omega_{k}\rangle+
g^{2(n+1)}\sum_{k=0}^{n}c_{k}^{(n)}g^{-2n-2k}\frac{1}{g^{4}}
\bigg[\langle F\Omega_{k+1}\rangle_{0}-\langle F\Omega_{k}\rangle_{0}\langle\Omega_{1}\rangle_{0}
\bigg]\notag \\
&= -\sum_{k=0}^{n}c_{k}^{(n)}(n+k)\langle F \Omega_{k}\rangle+
\sum_{k=0}^{n}c_{k}^{(n)}
\bigg[\langle F\Omega_{k+1}\rangle-\langle F\Omega_{k}\rangle\langle\Omega_{1}\rangle
\bigg]. 
\end{align}
This is true if 
\begin{align}
\norm{\Omega_{n+1}} &= \sum_{k=0}^{n}c_{k}^{(n)}\bigg[\Omega_{k+1}-(n+k+\alpha)\,\Omega_{k}\bigg].
\end{align}
Indeed,  replacing in the r.h.s. the expression of $c_{k}^{(n)}$ one finds the correct expression 
 $\norm{\Omega_{n+1}} = \sum_{k=0}^{n+1}c^{(n+1)}_{k}\Omega_{k}$. By induction, 
 (\ref{4.13})  is proved.

\section{Asymptotic expansion of the function $\log\mc F(\kappa)$}
\la{app:asy}

Let us expand the integral representation (\ref{6.7}) at large $\kappa>0$. We can split the leading piece by 
adding and subtracting it. This gives
\be
\la{C.1}
\log \mc F(\kappa) = -\frac{\log 2}{\pi^{2}}\,\kappa + 4\, \int_{0}^{\infty}\frac{dt}{t^{2}}\ 
\frac{J_{0}(\frac{t\sqrt\kappa}{\pi})+t\,\frac{\sqrt\kappa}{\pi}\,J_{1}(\frac{t\sqrt\kappa}{\pi})
-1}{e^{t}+1}.
\ee
The second piece is a Mellin convolution $\int_{0}^{\infty} dt\,h(\lambda\,t)\,f(t)$, where 
$\lambda = \frac{\sqrt\kappa}{\pi}$ and we are interested in
the asymptotic expansion for $\lambda\gg 1$. From the elementary Mellin transforms \footnote{As usual,
$\mc M[f(t)] = \int_{0}^{\infty}dt\,t^{s-1}\,f(t)$.}
\begin{align}
\mc M\bigg[\frac{1}{e^{t}+1}\bigg] &=2^{-s} (2^s-2) \zeta (s) \Gamma (s),\notag \\
\mc M\bigg[J_{0}(t)-1\bigg] &=\frac{2^{s-1} \Gamma (\frac{s}{2})}{\Gamma (1-\frac{s}{2})},\qquad
\mc M\bigg[J_{1}(t)\bigg] =\frac{2^{s-1} \Gamma (\frac{s}{2}+\frac{1}{2})}{\Gamma \
(\frac{3}{2}-\frac{s}{2})},
\end{align}
we derive by Mellin inversion the asymptotic expansion of the integral in (\ref{C.1}) 
by computing the residue
\be
\mathop{\text{Res}}_{s=0}\bigg[\lambda^{-s}\,
\frac{(2^{s+2}-1) \pi ^{-s} (s+1)  \zeta (s+2)}{(\cos \
(\pi  s)-1) \Gamma (1-\frac{s}{2})^2}
\bigg] = \log\lambda-\frac{7\,\log 2}{3}-12\,\zeta'(-1).
\ee
Adding the leading term, this gives
\be
\log\mc F(\kappa) = -\frac{\log 2}{\pi^{2}}\,\kappa+\frac{1}{2}\log\left(\frac{\kappa}{\pi^{2}}\right)
-\frac{7\,\log 2}{3}-12\,\zeta'(-1)+\dots,
\ee
where dots stand for contributions that vanish with $\kappa\to\infty$ faster than any inverse power of $\kappa$.
\footnote{To give an idea of how fast they vanish, 
their weight relative to the leading term is already about $10^{-15}$  at $\kappa=1000$.}

\bibliography{BT-Biblio}
\bibliographystyle{JHEP}

\end{document}